\documentclass[prl,aps,amsmath,amssymb,twocolumn,groupedaddress,floats,showpacs,final]{revtex4-1}
\usepackage{graphicx}
\usepackage{dcolumn}
\usepackage{bm}
\usepackage{color}
\usepackage{hyperref}
\hypersetup{
  pdftitle={Bold Diagrammatic Monte Carlo Method Applied to Fermionized Frustrated Spins},
  pdfauthor={S. A. Kulagin, N. Prokof'ev, O. A. Starykh, B. Svistunov, and C. N. Varney},
  colorlinks=true,
  linkcolor=blue,
  citecolor=blue,
  pdfpagemode=UseNone
}

\begin{document}

\title{Bold Diagrammatic Monte Carlo Method Applied to Fermionized Frustrated Spins}

\author{S.~A. Kulagin$^{1,2}$, N. Prokof'ev$^{1,3}$, O.~A. Starykh$^{4}$, B. Svistunov$^{1,3}$, and C.~N. Varney$^{1}$}

\affiliation{
$^1$Department of Physics, University of Massachusetts,
  Amherst, Massachusetts 01003, USA\\
$^2$Institute for Nuclear Research of Russian Academy of Sciences,
  117312 Moscow, Russia\\
$^3$Russian Research Center ``Kurchatov Institute",
  123182 Moscow, Russia\\
$^4$Department of Physics and Astronomy, University of
  Utah, Salt Lake City, Utah 84112, USA
}

\date{\today}
\begin{abstract}
  We demonstrate, by considering the triangular lattice spin-$1/2$
  Heisenberg model, that Monte Carlo sampling of skeleton Feynman
  diagrams within the fermionization framework offers a universal
  first-principles tool for strongly correlated lattice quantum
  systems.  We observe the fermionic sign blessing---cancellation of
  higher order diagrams leading to a finite convergence radius of the
  series. We calculate the magnetic susceptibility of the
  triangular-lattice quantum antiferromagnet in the correlated
  paramagnet regime and reveal a surprisingly accurate microscopic
  correspondence with its classical counterpart at all
  accessible temperatures. The extrapolation of the observed relation to zero temperature
  suggests the absence of the magnetic order in the ground state.
  We critically examine the implications of this unusual scenario.
\end{abstract}


\maketitle

%\begin{multicols}{2}
The method of bold diagrammatic Monte Carlo simulation (BDMC)~\cite{bold1} allows
one to sample contributions from millions of skeleton Feynman diagrams
and extrapolate the results to the infinite diagram order, provided the
series is convergent (or subject to resummation beyond the
convergence radius).  Recent experimentally certified application of
BDMC to unitary fermions down to the point of the superfluid
transition~\cite{NatureP} makes a strong case for BDMC method as a generic
method for dealing with correlated fermions described by Hamiltonians
without small parameters. One intriguing avenue to explore is to apply
it to frustrated lattice spin systems, where, on one hand, standard
Monte Carlo (MC) simulation fails because of the sign problem~\cite{Loh1990}, and, on
the other hand, the system's Hamiltonian can be always written in the
fermionic representation~\cite{PopovFedotov1,PopovFedotov2,
  Fermionization} which contains no large parameters---exactly what
is needed for the anticipated convergence of BDMC series with the diagram
order.

The BDMC approach is based on the {\it sign blessing} phenomenon,
when, despite the factorial increase in the number of diagrams with
expansion order, the series features a finite convergence radius
because of dramatic (sign alternation induced) compensation between
the diagrams. With the finite convergence radius, the series can be
summed either directly, or with resummation techniques that can be
potentially applied down to the critical temperature of the
phase transition, if any. (At the critical temperature thermodynamic functions
become nonanalytic, and the diagrammatic expansion involving explicit
symmetry breaking by the finite order parameter is necessary to treat the critical region and the
phase with broken symmetry.) In the absence of the sign blessing, the
resummation protocols become questionable in view of the known
mathematical theorems regarding asymptotic series. At the moment,
there is no theory allowing one to prove the existence of a finite
convergence radius analytically.
The absence of Dyson's collapse~\cite{Dyson} in a given fermionic system is merely providing hope
that the corresponding diagrammatic series is not asymptotic and
cannot be {\it a priori} taken as a sufficient condition for the sign
blessing. Hence, the applicability of BDMC method to a given system can be
established only on the basis of a direct numerical evidence for
series convergence and comparison with either experiment or
alternative controllable techniques, such as high-temperature series \cite{singh05} and
numerical linked cluster (NLC) expansions \cite{Rigol}.
In this Letter, we report the first successful application of BDMC method to
fermionized quantum spin systems by simulating the canonical model
of frustrated quantum magnetism---the triangular lattice
antiferromagnetic spin-1/2 Heisenberg model (TLHA).
We demonstrate that BDMC method for this frustrated magnet {\em is indeed} subject to
the sign blessing phenomenon which allows us to obtain basic static
and dynamic correlation functions with controllable (about one percent
or better) accuracy. The agreement with extrapolated high-temperature expansions
is excellent.

In addition, we report a very surprising finding of extreme similarity between short-distance static spin correlations of the
quantum and classical spin models, evaluated at different but uniquely related to each other temperatures.
This accurate (within the error bars) quantum-to-classical correspondence holds
at all temperatures accessible to us,
$T\geq 0.375 $ (here and below temperature is measured
in the units of the exchange constant $J$). Specifically, the {\it
  entire} static correlation function of the quantum model at a given
temperature $T$---having quite nontrivial pattern of sign-alternating
spatial dependence and temperature evolution, thus forming a system's
fingerprint---turns out to be equal, up to a global temperature
dependent normalization factor, to its classical counterpart at a
certain temperature $T_{\rm cl}\equiv T_{\rm cl}(T)$.
Extrapolation of the obtained $T_{\rm cl}(T)$ curve to the $T=0$ limit
results in a finite value of $T_{\rm cl}(0)>0$, suggesting
a quantum-disordered ground state of the quantum model.

The Hamiltonian of the TLHA is given by
\begin{equation}
H = J \sum_{<i,j>}  \, {\vec S}_i \cdot {\vec S}_j \, .
%\qquad (J>0) \, .
\label{HM}
\end{equation}
Here ${\vec S}_i$ is the spin-1/2 operator on the $i$th site of the
triangular lattice and the sum is over the nearest neighbor pairs coupled by the positive exchange integral, $J>0$. As
found by Popov and Fedotov~\cite{PopovFedotov1,PopovFedotov2}, the
grand canonical Gibbs distribution of the model \eqref{HM} can be
reformulated identically in terms of purely fermionic operators using
\begin{equation}
{\vec S}_i\,  =\,  \frac{1}{2}\sum_{\alpha, \beta}\,  f_{i \alpha}^{\dagger }
\vec{\sigma }_{\alpha \beta} f_{i \beta}^{\,}   \, ,
\label{SF}
\end{equation}
where $f_{i \beta}$ is the second quantized operator annihilating a
fermion with spin projection $\alpha,\beta =\pm 1$ on site $i$, and
$\vec{\sigma }$ are the Pauli matrices. The representation \eqref{SF}
leads to a flat-band fermionic Hamiltonian, $H_{\rm F}$, with two-body
interactions and amenable to direct diagrammatic treatment.  To
eliminate statistical contributions from nonphysical states having
either zero or two fermions, Ref.~\onlinecite{PopovFedotov1} introduced
an imaginary chemical potential to $H_{\rm F}$:
\begin{equation}
H_{\rm F} \to  H_{\rm F} -  i (\pi/2) T \sum_{i} (n_{i} -1) \, , \quad n_{i} =\sum_\alpha f_{i \alpha}^{\dagger }f_{i \alpha}^{\,}\, .
\label{HF}
\end{equation}
The added term commutes with the original Hamiltonian and has no
effect on properties of the physical subspace $\{ n_i=1 \}$
whatsoever. Moreover, the grand canonical partition functions and
spin-spin correlation functions of the original spin model and its
fermionic version are also identical because (i) physical and
nonphysical sites decouple in the trace and (ii) the trace over
nonphysical states yields identical zero on every site.  As a result,
one arrives at a rather standard Hamiltonian for fermions interacting
through two-body terms. A complex value of the chemical potential,
which can also be viewed as a peculiar shift of the fermonic Matsubara frequency
$\omega_n = 2\pi (n+1/2) T \to 2\pi (n+1/4) T$,
is a small price to pay for the luxury of having the
diagrammatic technique.

\begin{figure}[htbp]
\includegraphics[angle=0,width=0.8\columnwidth]{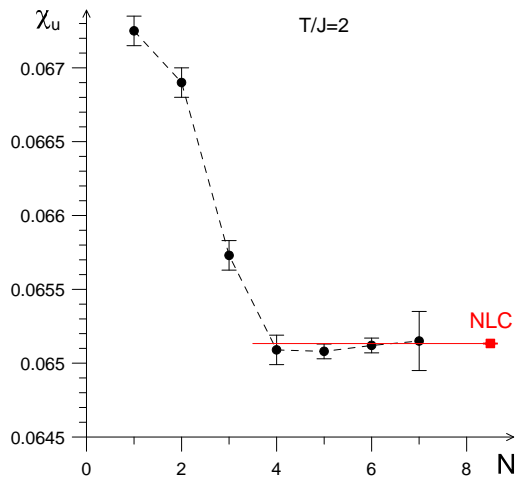}
\caption{\label{fig:1} (Color online) Uniform susceptibility
  calculated within the $G^2W$-skeleton expansion as a function of the
  maximum diagram order retained in the BDMC simulation (black dots)
  for $T/J=2$. The result of the high-temperature expansion (with Pad\'e
  approximant extrapolation)~\cite{singh05,Rigol} is shown by the red square and horizontal
  line.  }
\end{figure}

We perform BDMC simulations using the standard
$G^2W$-skeleton diagrammatic expansion of the fermionic model
\eqref{HM}-\eqref{HF} in the real space--imaginary time
representation~\cite{kpssv2013}, see also \cite{Kris08}.
The first and most important question to answer is
whether the sign blessing phenomenon indeed takes place. In
Fig.~\ref{fig:1} we show comparison between the calculated answer for
the static uniform magnetic susceptibility, $\chi_u$, and the NLC
expansion result~\cite{Rigol} at $T=2$. This temperature is low enough
to ensure that we are in the regime of strong correlations because
$\chi_u$ is nearly a factor of two smaller than the free-spin answer
$\chi_u^{(0)} = 1/4T$.  On the other hand, this temperature is high
enough to be sure that the high-temperature series can be described by Pad\'e
approximants without significant systematic deviations from the exact
answer~\cite{singh05,Rigol} (at slightly lower temperature the bare NLC series
starts to diverge).  We clearly see that the BDMC series converges to
the correct result with an accuracy of about three meaningful digits and
there is no statistically significant change when more than a hundred
thousand of 7th order diagrams~\cite{remark2} are accounted for.  The
error bar for the 7th order point is significantly increased due to
factorial growth in computational complexity. 
Feynman diagrams are usually formulated for the system in the thermodynamic limit.
In practice, for reasons of convenient data handling, our code works with 
finite system sizes $L$ with periodic boundary conditions 
(its performance does not depend on $L$). In all cases we choose $L$ to 
be much larger than the correlation length $\xi$ and check that doubling 
the system size makes no detectable changes in the final answer.
The 4th order result can be obtained after several hours of CPU time on a single processor.

\begin{figure}[htbp]
\includegraphics[angle=0,width=0.8\columnwidth]{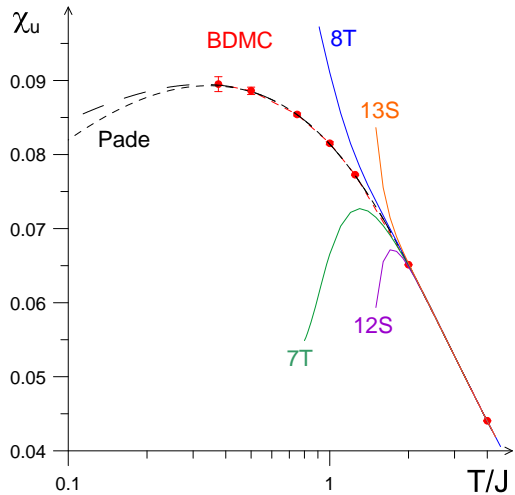}
\caption{\label{fig:2} (Color online) Uniform susceptibility as a
  function of temperature (red dots) for the triangular Heisenberg
  antiferromagnet calculated within the BDMC approach.  NLC expansion
  results based on triangles (labeled as 7T and 8T) and sites (labeled
  as 12S and 13S)~\cite{Rigol} are shown along with two different Pad\'e
  approximant extrapolations~\cite{singh05}.  }
\end{figure}

Interestingly enough, when temperature is lowered down to $T=1$ which
is significantly below the point where the bare NLC series start to
diverge, see Fig.~\ref{fig:2}, the BDMC series continue to converge
exponentially.  This underlines the importance of performing
simulations within the self-consistent skeleton formulation.
%
%%%%%%%%%%%%%%%%%%%%%%%%%%%%%%%%%%%%%%%%%%%%%%%%%%%%%%%%%%%%%%
%%%%%%%%%%%%%%%%%%%%%%%%%%%%%%%%%%%%%%%%%%%%%%%%%%%%%%%%%%%%%%

In Fig.~\ref{fig:2} we show results of the BDMC simulation performed
at temperatures significantly below the mean-field transition
temperature. We observe excellent agreement (within our error bars)
with the Pad\'e approximants used to extrapolate the high-temperature series
data to lower temperature~\cite{singh05}.  Within the current protocol of dealing with skeleton
diagrams we were not able to go to a lower temperature due to the
development of near singularity in the response function (and thus in
the effective-interaction propagator) at the classical ordering wave vector ${\mathbf
  Q}=(4\pi/3,0)$, in units of inverse lattice constant.
In future work, we plan to apply pole-regularization
schemes to overcome this technical problem.

%
%%%%%%%%%%%%%%%%%%%%%%%%%%%%%%%%%%%%%%%%%%%%%%%%%%%%%%%%%%%%%%
%%%%%%%%%%%%%%%%%%%%%%%%%%%%%%%%%%%%%%%%%%%%%%%%%%%%%%%%%%%%%%

\begin{figure*}
\hspace*{-0.6cm} \includegraphics[angle=0,width=0.75\columnwidth]{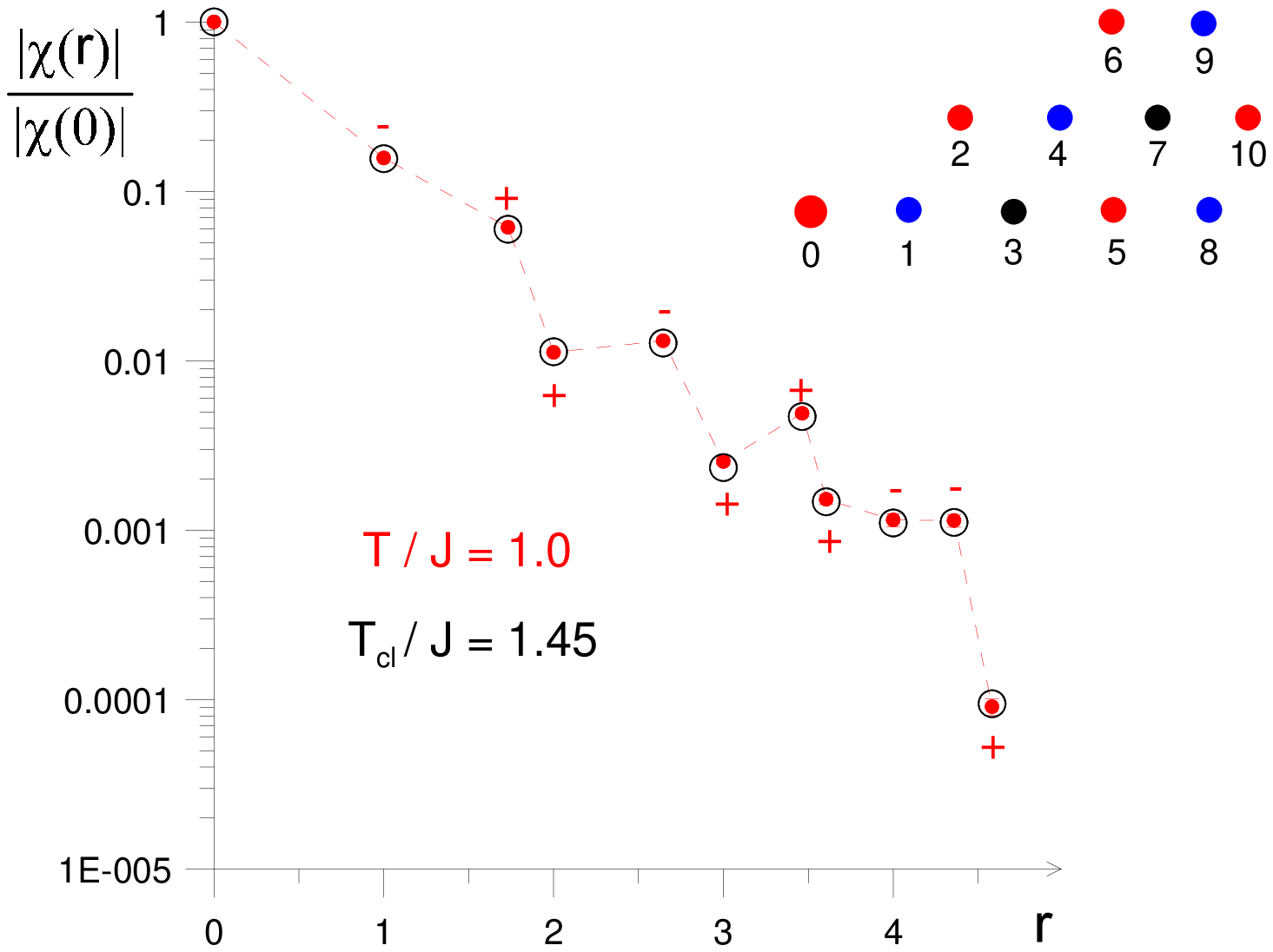}
\includegraphics[angle=0,width=0.63\columnwidth]{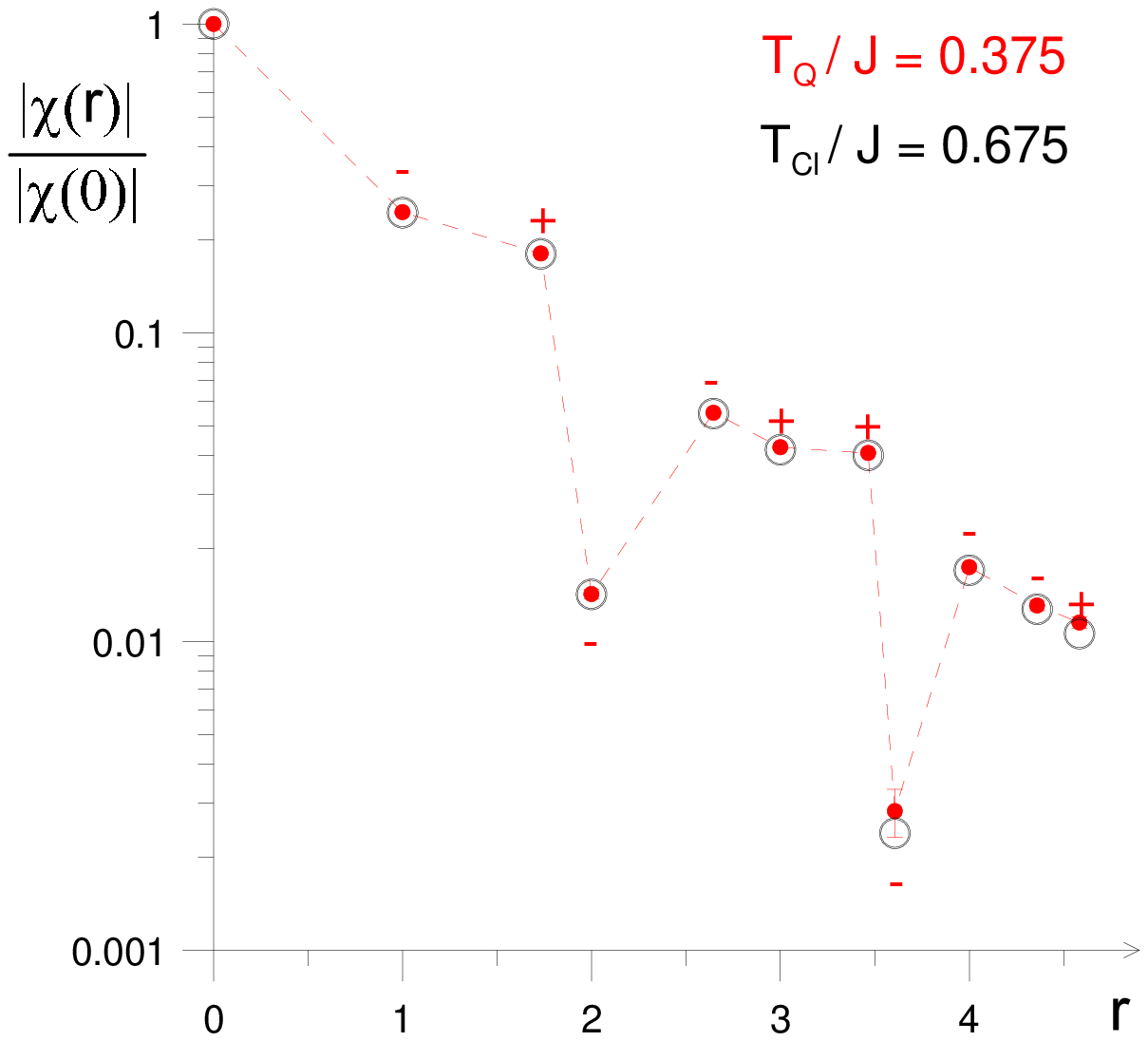}
\includegraphics[angle=0,width=0.7\columnwidth]{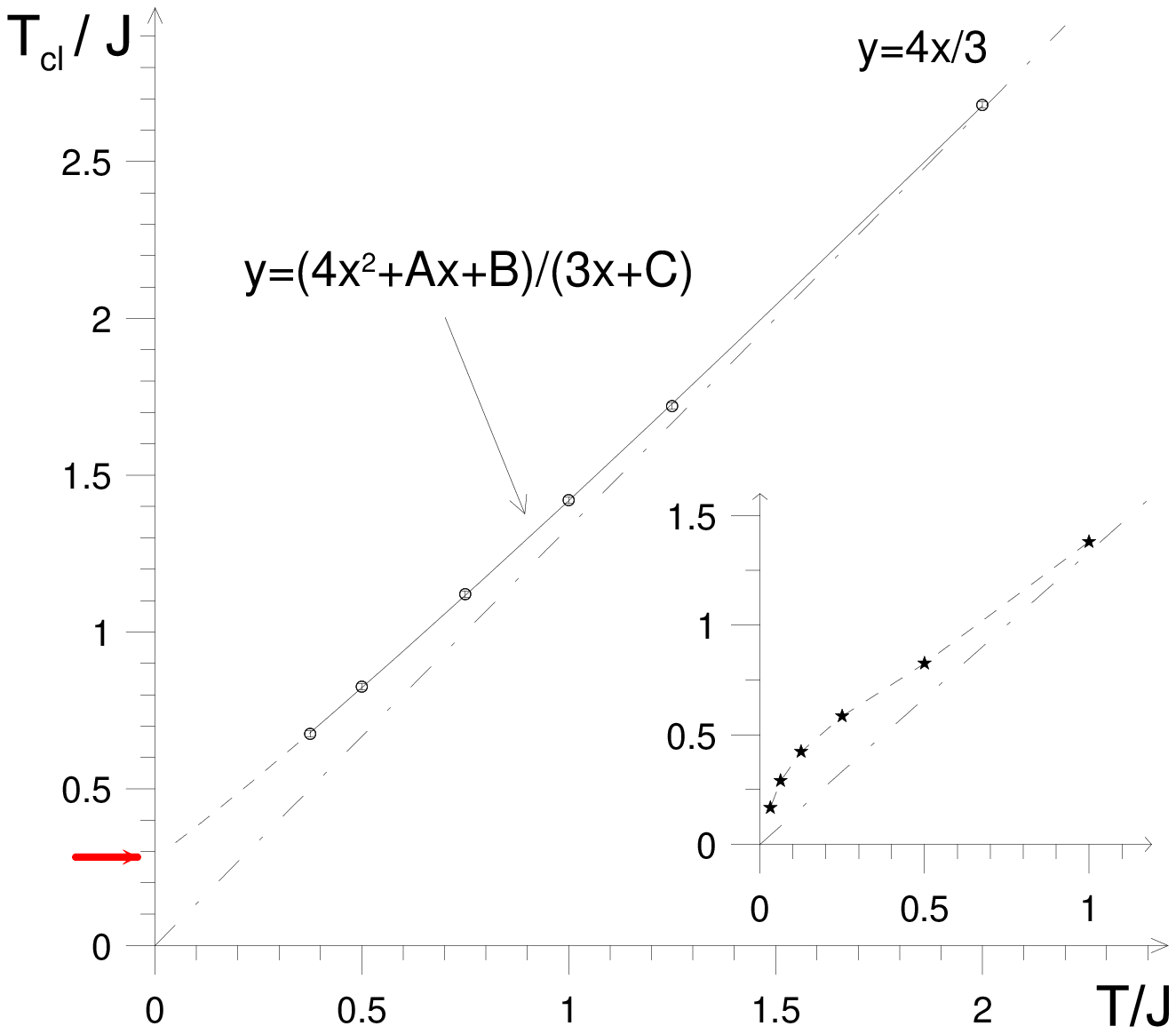}
\caption{
  (Color online) Left and middle panels: Perfect match of the normalized quantum and classical
  responses at the corresponding temperatures, $T$ and
  $T_{\rm cl}(T)$. Points are ordered according to their distance from the
  origin, $r$, as is illustrated in the right top corner. The sign of
  the correlator is indicated explicitly next to the point. Right Panel:
  Mapping between the classical and quantum
  temperatures. The solid (continued as dashed) line is the fitting function,
  $y=(4x^2+Ax+B)/(3x+C)$, with $A=0.462$, $B=1.065$, $C=3.825$, which
  satisfies the asymptotic law $y=4x/3$ expected in the
  high-temperature limit shown by the dash-dotted line.
  The red arrow indicates  the position of the chiral transition advocated in
  Refs.~\cite{kawamura84,kawamura10,xu95}. For comparison,
  the square lattice data (stars) are shown in the inset.
  \label{fig:fingerprint}
}
\end{figure*}

We now turn to the static susceptibility
\begin{equation}
\chi({\bf r}) \, =\, \int_0^{1/T} d\tau \, \langle\,  S^z_{\bf 0}(0)
\,  S^z_{\bf r} (\tau) \, \rangle \, .
\label{K}
\end{equation}
Here $S^z_{\bf r}(\tau)$ is the Matsubara spin operator on the lattice
site labeled by the integer index vector ${\bf r}$.  For the simplicity of
comparing susceptibility \eqref{K} with its classical counterpart, we
normalize it to unity at the origin, $\chi({\bf r}) \to \chi({\bf r})/
\chi({\bf 0})$, doing the same with the classical $\chi_{\rm cl}({\bf r})$. The
latter is obtained by Metropolis simulation of the classical
Heisenberg model \eqref{HM} in which quantum
spin operators are replaced with classical unit vectors ${\bf n}_{\bf r}$.  For every
accessible temperature $T$ we observe a perfect match (within the error bars, which are about $1\%$)
between quantum correlator $\chi({\bf r})$ and its classical counterpart, for $r$ ranging from 1 to 5 (which includes 10 different sites),
calculated at a certain temperature $T_{\rm cl}(T)$.
A typical example of the match is presented in Fig.~\ref{fig:fingerprint}. We note in
passing that the equal-time correlation function, $ \langle\, S^z_{\bf
  0}(0) \, S^z_{\bf r} (0) \, \rangle$, while having qualitatively
similar shape to that of \eqref{K}, does not match the classical
correlator $\chi_{\rm cl}({\bf r}) = \langle\, n^z_{\bf 0} \, n^z_{\bf r} \, \rangle $,
especially so for sites at which the sign of the correlation
changes with temperature (such as sites 3 and 7 in Fig.~\ref{fig:fingerprint}
for which the sign of $\chi({\bf r})$ changes
from ferromagnetic at high $T$ to antiferromagnetic one below $T \approx 0.5$).

Mapping long-range correlations in quantum models onto the
renormalized classical behavior is rather standard on approach to the
ordered state \cite{sachdev_book}. What we observe is fundamentally different:
quantum-to-classical correspondence, or QCC, is valid in the
intermediate temperature regime at all distances, including
nearest-neighbor sites, and when the correlation length $\xi$ is still very short,
of the order of the lattice spacing, $\xi \sim 1$.
It is worth noting that this short-distance correspondence is also very different from
the high-$T$ quasiclassical wave regime of Ref.~\onlinecite{sachdev_book}
which allows for the classical description at distances $r \sim \xi \gg 1$.

We find that QCC also takes place for the $s=1/2$ square lattice
Heisenberg antiferromagnet where, thanks to the absence of a sign problem for the path-integral
Monte Carlo simulation, it has a relative accuracy of $\sim 0.3\%$ at all temperatures 
(down to the ground state in a finite-size system).
These facts suggest that QCC in 2D is extremely accurate  
and thus may take place for other lattices
(however, it does not hold for 1D chains).

%%%%%%%%%%%%%%%%%%%%%%%%%%%%%%%%%%%%%%%%%%%%%%%%%%%%%%%%%%%%%
%%%%%%%%%%%%%%%%%%%%%%%%%%%%%%%%%%%%%%%%%%%%%%%%%%%%%%%%%%%%%%%%%

The quality of the matching procedure allows us to establish the
$T_{\rm cl}(T)$ correspondence with an accuracy of about one percent.
In the right panel of Fig.~\ref{fig:fingerprint} we plot the final result along with the
asymptotic high-temperature relation $T_{\rm cl}=(4/3)T$ reflecting
the difference between the $(S^z)^2=1/4$ and $\langle (n^z)^2 \rangle =
1/3$. An immediate consequence of observed QCC in Fig.~\ref{fig:fingerprint}
is that the entire $\bf{q}$ dependence of the static susceptibility $\chi({\bf q}, \omega_n=0)$
of the quantum model
is given by the susceptibility of the classical model at temperature $T_{\rm cl}(T)$,
which is readily available from classical Monte Carlo simulations.

Due to the limited low-temperature range of the $T_{\rm cl}(T)$ curve for the TLHA
it is perhaps too early to make any definite conclusion regarding its extrapolation
down to the $T=0$ limit. One possibility is that it smoothly extrapolates to a finite value
$T_{\rm cl}(0)=0.28$, implying that the ground state is some kind of a spin liquid.
This possibility was discussed by Anderson~\cite{Anderson73} almost forty years ago
but was subsequently rejected on the basis of numerous investigations which include exact
diagonalization~\cite{bernu92,sindzingre,yunoki}, Green's function MC calculations~\cite{capriotti},
series expansion \cite{zheng1999}, density matrix renormalization
group ~\cite{white07} studies, as well as large-$S$ (spin wave) \cite{miyake1992,chubukov1994,chernyshev2009},
large-N \cite{sachdev1992},
and functional renormalization group analysis \cite{thomale2011}.
Note, however, that the spin correlation length for the classical model at $T_{\rm cl} \approx 0.28$
is above $10^3$ lattice periods~\cite{kawamura10} and thus simulations of small system sizes
$L\sim 10$ would be severely affected by finite-size effects.
The value of $T_{\rm cl}(0) \approx 0.28 $ is surprisingly close
(essentially within the error bars) to the temperature obtained by
extrapolating transition temperatures for the $q=3$ Potts transition
in finite magnetic fields $h$ to the $h=0$ limit~\cite{misha,seabra}.
Large-scale MC simulations performed in zero magnetic field
also identify $T_{\rm cl}=0.285(5)$ as the critical point of the
chiral transition~\cite{kawamura84,kawamura10,xu95}.
However, the debate
with regards to the existence of the chiral transition is not settled
yet---an alternative scenario \cite{azaria92,southern93} predicts
a sharp crossover to a more standard nonlinear sigma-model type behavior around $T_{\rm cl}=0.28$.

%
%%%%%%%%%%%%%%%%%%%%%%%%%%%%%%%%%%%%%%%%%%%%%%%%%%%%%%%%%%%%%%
\begin{figure}[tbp]
\includegraphics[angle=0,width=0.8\columnwidth]{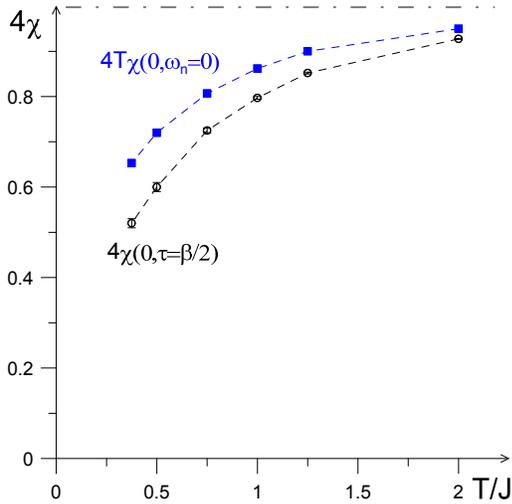}
\caption{\label{fig:6}
  (Color online) Blue (square) symbols: local static susceptibility $\chi(r=0,\omega_m=0)$, multiplied by $4 T$, as a function of $T/J$.
  Black circles show $T$ dependence of the ($4$ times) local spin correlation function $\chi(r=0,\tau=\beta/2)$.
}
\end{figure}
%%%%%%%%%%%%%%%%%%%%%%%%%%%%%%%%%%%%%%%%%%%%%%%%%%%%%%%%%%%%%%
%

The other possibility is that the QCC curve $T_{\rm cl}(T)$ will cross over to the standard
renormalized classical behavior in the long wavelength limit and will arrive at $T_{\rm cl}(0) = 0$,
implying the ordered quantum ground state. This is exactly what happens for the square lattice antiferomagnet,
see inset in the right panel in Fig.~\ref{fig:fingerprint}.
In fact, it is also known (and can be readily deduced from the correspondence plot and Fig.2 of Ref.~\onlinecite{kawamura10})
that in the TLHA the renormalized classical regime with large correlation length emerges only below temperature
$T \approx 0.25$ \cite{singh,zheng2006}, which is well below our lowest data point $T=0.375$.
Clearly, more data at lower temperatures are required in order to resolve this fascinating question.

The normalization factor $\chi(0) \equiv \chi(r=0,\omega_m=0)$ in Fig.~\ref{fig:fingerprint} is given
by the local static susceptibility which of course is different for the classical and quantum system.
For the classical Heisenberg model $T\chi(0)$ is simply $1/3$, independent of temperature, while in the quantum
system this quantity is $T$ dependent, as Fig.~\ref{fig:6} shows.
The same figure also shows the local spin correlation function at
$\tau=\beta/2$, $\chi(r=0,\tau=\beta/2) = T \sum_m e^{i \pi m} \chi(r=0,\omega_m)$.
This too probes quantum fluctuations, i.e.,
contributions to the sum from terms with $\omega_m \neq 0$.
As expected, both curves deviate from unity with the lowering of $T$,
reflecting the increasing role of quantum fluctuations.

Perhaps the most striking feature of QCC is its predictive power in the search for spin-liquid states.
Indeed, if QCC is confirmed for a given model of quantum magnetism and the classical ground state
is not ordered due to macroscopic degeneracy then the quantum ground state is not ordered as well;
i.e., it is a spin liquid. 
Moreover, even if the classical ground state is ordered but the correspondence
curve $T_{\rm cl}(T)$ is such that $T_{\rm cl}(0)\ne 0$, the quantum ground state is still not ordered
similarly to its finite-temperature classical counterpart.
While the final outcome for the TLHA remains to be seen, our data convincingly show an
unusual classical-to-quantum correspondence with regards to the static spin correlations.

We thank M. Rigol for comments and the original NLC data and R.R.P. Singh for comments.
This work was supported by the National Science Foundation
under Grants No. PHY-1005543 (S.K., N.P., B.S., and C.N.V.)
and No. DMR-1206774 (O.A.S.), and by a grant from the Army Research Office
with funding from the DARPA.

\bibliography{references}

\begin{thebibliography}{35}%
\makeatletter
\providecommand \@ifxundefined [1]{%
 \@ifx{#1\undefined}
}%
\providecommand \@ifnum [1]{%
 \ifnum #1\expandafter \@firstoftwo
 \else \expandafter \@secondoftwo
 \fi
}%
\providecommand \@ifx [1]{%
 \ifx #1\expandafter \@firstoftwo
 \else \expandafter \@secondoftwo
 \fi
}%
\providecommand \natexlab [1]{#1}%
\providecommand \enquote  [1]{``#1''}%
\providecommand \bibnamefont  [1]{#1}%
\providecommand \bibfnamefont [1]{#1}%
\providecommand \citenamefont [1]{#1}%
\providecommand \href@noop [0]{\@secondoftwo}%
\providecommand \href [0]{\begingroup \@sanitize@url \@href}%
\providecommand \@href[1]{\@@startlink{#1}\@@href}%
\providecommand \@@href[1]{\endgroup#1\@@endlink}%
\providecommand \@sanitize@url [0]{\catcode `\\12\catcode `\$12\catcode
  `\&12\catcode `\#12\catcode `\^12\catcode `\_12\catcode `\%12\relax}%
\providecommand \@@startlink[1]{}%
\providecommand \@@endlink[0]{}%
\providecommand \url  [0]{\begingroup\@sanitize@url \@url }%
\providecommand \@url [1]{\endgroup\@href {#1}{\urlprefix }}%
\providecommand \urlprefix  [0]{URL }%
\providecommand \Eprint [0]{\href }%
\providecommand \doibase [0]{http://dx.doi.org/}%
\providecommand \selectlanguage [0]{\@gobble}%
\providecommand \bibinfo  [0]{\@secondoftwo}%
\providecommand \bibfield  [0]{\@secondoftwo}%
\providecommand \translation [1]{[#1]}%
\providecommand \BibitemOpen [0]{}%
\providecommand \bibitemStop [0]{}%
\providecommand \bibitemNoStop [0]{.\EOS\space}%
\providecommand \EOS [0]{\spacefactor3000\relax}%
\providecommand \BibitemShut  [1]{\csname bibitem#1\endcsname}%
\let\auto@bib@innerbib\@empty
%</preamble>
\bibitem [{\citenamefont {Prokof'ev}\ and\ \citenamefont
  {Svistunov}(2007)}]{bold1}%
  \BibitemOpen
  \bibfield  {author} {\bibinfo {author} {\bibfnamefont {N.}~\bibnamefont
  {Prokof'ev}}\ and\ \bibinfo {author} {\bibfnamefont {B.}~\bibnamefont
  {Svistunov}},\ }\href {\doibase 10.1103/PhysRevLett.99.250201} {\bibfield
  {journal} {\bibinfo  {journal} {Phys. Rev. Lett.}\ }\textbf {\bibinfo
  {volume} {99}},\ \bibinfo {pages} {250201} (\bibinfo {year}
  {2007})}\BibitemShut {NoStop}%
\bibitem [{\citenamefont {{Van Houcke}}\ \emph {et~al.}(2012)\citenamefont
  {{Van Houcke}}, \citenamefont {Werner}, \citenamefont {Kozik}, \citenamefont
  {Prokof'ev}, \citenamefont {Svistunov}, \citenamefont {Ku}, \citenamefont
  {Sommer}, \citenamefont {Cheuk}, \citenamefont {Schirotzek},\ and\
  \citenamefont {Zwierlein}}]{NatureP}%
  \BibitemOpen
  \bibfield  {author} {\bibinfo {author} {\bibfnamefont {K.}~\bibnamefont {{Van
  Houcke}}}, \bibinfo {author} {\bibfnamefont {F.}~\bibnamefont {Werner}},
  \bibinfo {author} {\bibfnamefont {E.}~\bibnamefont {Kozik}}, \bibinfo
  {author} {\bibfnamefont {N.}~\bibnamefont {Prokof'ev}}, \bibinfo {author}
  {\bibfnamefont {B.}~\bibnamefont {Svistunov}}, \bibinfo {author}
  {\bibfnamefont {M.~J.~H.}\ \bibnamefont {Ku}}, \bibinfo {author}
  {\bibfnamefont {A.~T.}\ \bibnamefont {Sommer}}, \bibinfo {author}
  {\bibfnamefont {L.~W.}\ \bibnamefont {Cheuk}}, \bibinfo {author}
  {\bibfnamefont {A.}~\bibnamefont {Schirotzek}}, \ and\ \bibinfo {author}
  {\bibfnamefont {M.~W.}\ \bibnamefont {Zwierlein}},\ }\href {\doibase
  10.1038/nphys2273} {\bibfield  {journal} {\bibinfo  {journal} {Nat. Phys.}\
  }\textbf {\bibinfo {volume} {8}},\ \bibinfo {pages} {366} (\bibinfo {year}
  {2012})}\BibitemShut {NoStop}%
\bibitem [{\citenamefont {Loh}\ \emph {et~al.}(1990)\citenamefont {Loh},
  \citenamefont {Gubernatis}, \citenamefont {Scalettar}, \citenamefont {White},
  \citenamefont {Scalapino},\ and\ \citenamefont {Sugar}}]{Loh1990}%
  \BibitemOpen
  \bibfield  {author} {\bibinfo {author} {\bibfnamefont {E.~Y.}\ \bibnamefont
  {Loh}}, \bibinfo {author} {\bibfnamefont {J.~E.}\ \bibnamefont {Gubernatis}},
  \bibinfo {author} {\bibfnamefont {R.~T.}\ \bibnamefont {Scalettar}}, \bibinfo
  {author} {\bibfnamefont {S.~R.}\ \bibnamefont {White}}, \bibinfo {author}
  {\bibfnamefont {D.~J.}\ \bibnamefont {Scalapino}}, \ and\ \bibinfo {author}
  {\bibfnamefont {R.~L.}\ \bibnamefont {Sugar}},\ }\href {\doibase
  10.1103/PhysRevB.41.9301} {\bibfield  {journal} {\bibinfo  {journal} {Phys.
  Rev. B}\ }\textbf {\bibinfo {volume} {41}},\ \bibinfo {pages} {9301}
  (\bibinfo {year} {1990})}\BibitemShut {NoStop}%
\bibitem [{\citenamefont {Popov}\ and\ \citenamefont
  {Fedotov}(1988)}]{PopovFedotov1}%
  \BibitemOpen
  \bibfield  {author} {\bibinfo {author} {\bibfnamefont {V.~N.}\ \bibnamefont
  {Popov}}\ and\ \bibinfo {author} {\bibfnamefont {S.~A.}\ \bibnamefont
  {Fedotov}},\ }\href
  {http://www.jetp.ac.ru/cgi-bin/e/index/e/67/3/p535?a=list} {\bibfield
  {journal} {\bibinfo  {journal} {Sov. Phys. - JETP}\ }\textbf {\bibinfo
  {volume} {67}},\ \bibinfo {pages} {535} (\bibinfo {year} {1988})}\BibitemShut
  {NoStop}%
\bibitem [{\citenamefont {Popov}\ and\ \citenamefont
  {Fedotov}(1991)}]{PopovFedotov2}%
  \BibitemOpen
  \bibfield  {author} {\bibinfo {author} {\bibfnamefont {V.~N.}\ \bibnamefont
  {Popov}}\ and\ \bibinfo {author} {\bibfnamefont {S.~A.}\ \bibnamefont
  {Fedotov}},\ }\href@noop {} {\bibfield  {journal} {\bibinfo  {journal} {Proc.
  Steklov Inst. Math.}\ }\textbf {\bibinfo {volume} {177}},\ \bibinfo {pages}
  {184} (\bibinfo {year} {1991})}\BibitemShut {NoStop}%
\bibitem [{\citenamefont {Prokof'ev}\ and\ \citenamefont
  {Svistunov}(2011)}]{Fermionization}%
  \BibitemOpen
  \bibfield  {author} {\bibinfo {author} {\bibfnamefont {N.~V.}\ \bibnamefont
  {Prokof'ev}}\ and\ \bibinfo {author} {\bibfnamefont {B.~V.}\ \bibnamefont
  {Svistunov}},\ }\href {\doibase 10.1103/PhysRevB.84.073102} {\bibfield
  {journal} {\bibinfo  {journal} {Phys. Rev. B}\ }\textbf {\bibinfo {volume}
  {84}},\ \bibinfo {pages} {073102} (\bibinfo {year} {2011})}\BibitemShut
  {NoStop}%
\bibitem [{\citenamefont {Dyson}(1952)}]{Dyson}%
  \BibitemOpen
  \bibfield  {author} {\bibinfo {author} {\bibfnamefont {F.~J.}\ \bibnamefont
  {Dyson}},\ }\href {\doibase 10.1103/PhysRev.85.631} {\bibfield  {journal}
  {\bibinfo  {journal} {Phys. Rev.}\ }\textbf {\bibinfo {volume} {85}},\
  \bibinfo {pages} {631} (\bibinfo {year} {1952})}\BibitemShut {NoStop}%
\bibitem [{\citenamefont {Zheng}\ \emph {et~al.}(2005)\citenamefont {Zheng},
  \citenamefont {Singh}, \citenamefont {McKenzie},\ and\ \citenamefont
  {Coldea}}]{singh05}%
  \BibitemOpen
  \bibfield  {author} {\bibinfo {author} {\bibfnamefont {W.}~\bibnamefont
  {Zheng}}, \bibinfo {author} {\bibfnamefont {R.~R.~P.}\ \bibnamefont {Singh}},
  \bibinfo {author} {\bibfnamefont {R.~H.}\ \bibnamefont {McKenzie}}, \ and\
  \bibinfo {author} {\bibfnamefont {R.}~\bibnamefont {Coldea}},\ }\href
  {\doibase 10.1103/PhysRevB.71.134422} {\bibfield  {journal} {\bibinfo
  {journal} {Phys. Rev. B}\ }\textbf {\bibinfo {volume} {71}},\ \bibinfo
  {pages} {134422} (\bibinfo {year} {2005})}\BibitemShut {NoStop}%
\bibitem [{\citenamefont {Rigol}\ \emph {et~al.}(2007)\citenamefont {Rigol},
  \citenamefont {Bryant},\ and\ \citenamefont {Singh}}]{Rigol}%
  \BibitemOpen
  \bibfield  {author} {\bibinfo {author} {\bibfnamefont {M.}~\bibnamefont
  {Rigol}}, \bibinfo {author} {\bibfnamefont {T.}~\bibnamefont {Bryant}}, \
  and\ \bibinfo {author} {\bibfnamefont {R.~R.~P.}\ \bibnamefont {Singh}},\
  }\href {\doibase 10.1103/PhysRevE.75.061118} {\bibfield  {journal} {\bibinfo
  {journal} {Phys. Rev. E}\ }\textbf {\bibinfo {volume} {75}},\ \bibinfo
  {pages} {061118} (\bibinfo {year} {2007})}\BibitemShut {NoStop}%
\bibitem [{\citenamefont {Kulagin}\ \emph {et~al.}(2013)\citenamefont
  {Kulagin}, \citenamefont {Prokof'ev}, \citenamefont {Starykh}, \citenamefont
  {Svistunov},\ and\ \citenamefont {Varney}}]{kpssv2013}%
  \BibitemOpen
  \bibfield  {author} {\bibinfo {author} {\bibfnamefont {S.}~\bibnamefont
  {Kulagin}}, \bibinfo {author} {\bibfnamefont {N.}~\bibnamefont {Prokof'ev}},
  \bibinfo {author} {\bibfnamefont {O.}~\bibnamefont {Starykh}}, \bibinfo
  {author} {\bibfnamefont {B.}~\bibnamefont {Svistunov}}, \ and\ \bibinfo
  {author} {\bibfnamefont {C.}~\bibnamefont {Varney}},\ }\href {\doibase
  10.1103/PhysRevB.87.024407} {\bibfield  {journal} {\bibinfo  {journal} {Phys.
  Rev. B}\ }\textbf {\bibinfo {volume} {87}},\ \bibinfo {pages} {024407}
  (\bibinfo {year} {2013})}\BibitemShut {NoStop}%
\bibitem [{\citenamefont {{Van Houcke}}\ \emph {et~al.}(2008)\citenamefont
  {{Van Houcke}}, \citenamefont {Kozik}, \citenamefont {Prokof'ev},\ and\
  \citenamefont {Svistunov}}]{Kris08}%
  \BibitemOpen
  \bibfield  {author} {\bibinfo {author} {\bibfnamefont {K.}~\bibnamefont {{Van
  Houcke}}}, \bibinfo {author} {\bibfnamefont {E.}~\bibnamefont {Kozik}},
  \bibinfo {author} {\bibfnamefont {N.}~\bibnamefont {Prokof'ev}}, \ and\
  \bibinfo {author} {\bibfnamefont {B.}~\bibnamefont {Svistunov}},\ }\enquote
  {\bibinfo {title} {{Diagrammatic Monte Carlo}},}\ in\ \href@noop {} {\emph
  {\bibinfo {booktitle} {Computer Simulation Studies in Condensed Matter
  Physics XXI}}},\ \bibinfo {editor} {edited by\ \bibinfo {editor}
  {\bibfnamefont {D.}~\bibnamefont {Landau}}, \bibinfo {editor} {\bibfnamefont
  {S.}~\bibnamefont {Lewis}}, \ and\ \bibinfo {editor} {\bibfnamefont
  {H.}~\bibnamefont {Schuttler}}}\ (\bibinfo  {publisher} {Springer Verlag},\
  \bibinfo {address} {Heidelberg, Berlin},\ \bibinfo {year} {2008})\BibitemShut
  {NoStop}%
\bibitem [{rem()}]{remark2}%
  \BibitemOpen
  \href@noop {} {}\bibinfo {note} {The number of topologically distinct
  diagrams within the $G^2W$-skeleton scheme was calculated in
  Ref.~\protect\cite{molinari}; the eight lowest orders are $1$, $1$, $6$,
  $49$, $542$, $7278$, $113824$, $2017881$}\BibitemShut {NoStop}%
\bibitem [{\citenamefont {Sachdev}(1999)}]{sachdev_book}%
  \BibitemOpen
  \bibfield  {author} {\bibinfo {author} {\bibfnamefont {S.}~\bibnamefont
  {Sachdev}},\ }\href@noop {} {\emph {\bibinfo {title} {Quantum Phase
  Transitions}}}\ (\bibinfo  {publisher} {Cambridge University Press},\
  \bibinfo {address} {Cambridge, England},\ \bibinfo {year} {1999})\BibitemShut
  {NoStop}%
\bibitem [{\citenamefont {Kawamura}\ and\ \citenamefont
  {Miyashita}(1984)}]{kawamura84}%
  \BibitemOpen
  \bibfield  {author} {\bibinfo {author} {\bibfnamefont {H.}~\bibnamefont
  {Kawamura}}\ and\ \bibinfo {author} {\bibfnamefont {S.}~\bibnamefont
  {Miyashita}},\ }\href {\doibase 10.1143/JPSJ.53.4138} {\bibfield  {journal}
  {\bibinfo  {journal} {J. Phys. Soc. Jpn.}\ }\textbf {\bibinfo {volume}
  {53}},\ \bibinfo {pages} {4138} (\bibinfo {year} {1984})}\BibitemShut
  {NoStop}%
\bibitem [{\citenamefont {Kawamura}\ \emph {et~al.}(2010)\citenamefont
  {Kawamura}, \citenamefont {Yamamoto},\ and\ \citenamefont
  {Okubo}}]{kawamura10}%
  \BibitemOpen
  \bibfield  {author} {\bibinfo {author} {\bibfnamefont {H.}~\bibnamefont
  {Kawamura}}, \bibinfo {author} {\bibfnamefont {A.}~\bibnamefont {Yamamoto}},
  \ and\ \bibinfo {author} {\bibfnamefont {T.}~\bibnamefont {Okubo}},\ }\href
  {\doibase 10.1143/JPSJ.79.023701} {\bibfield  {journal} {\bibinfo  {journal}
  {J. Phys. Soc. Jpn.}\ }\textbf {\bibinfo {volume} {79}},\ \bibinfo {pages}
  {023701} (\bibinfo {year} {2010})}\BibitemShut {NoStop}%
\bibitem [{\citenamefont {Southern}\ and\ \citenamefont {Xu}(1995)}]{xu95}%
  \BibitemOpen
  \bibfield  {author} {\bibinfo {author} {\bibfnamefont {B.}~\bibnamefont
  {Southern}}\ and\ \bibinfo {author} {\bibfnamefont {H.-J.}\ \bibnamefont
  {Xu}},\ }\href {\doibase 10.1103/PhysRevB.52.R3836} {\bibfield  {journal}
  {\bibinfo  {journal} {Phys. Rev. B}\ }\textbf {\bibinfo {volume} {52}},\
  \bibinfo {pages} {R3836} (\bibinfo {year} {1995})}\BibitemShut {NoStop}%
\bibitem [{\citenamefont {Anderson}(1973)}]{Anderson73}%
  \BibitemOpen
  \bibfield  {author} {\bibinfo {author} {\bibfnamefont {P.}~\bibnamefont
  {Anderson}},\ }\href {\doibase 10.1016/0025-5408(73)90167-0} {\bibfield
  {journal} {\bibinfo  {journal} {Materials Research Bulletin}\ }\textbf
  {\bibinfo {volume} {8}},\ \bibinfo {pages} {153 } (\bibinfo {year}
  {1973})}\BibitemShut {NoStop}%
\bibitem [{\citenamefont {Bernu}\ \emph {et~al.}(1992)\citenamefont {Bernu},
  \citenamefont {Lhuillier},\ and\ \citenamefont {Pierre}}]{bernu92}%
  \BibitemOpen
  \bibfield  {author} {\bibinfo {author} {\bibfnamefont {B.}~\bibnamefont
  {Bernu}}, \bibinfo {author} {\bibfnamefont {C.}~\bibnamefont {Lhuillier}}, \
  and\ \bibinfo {author} {\bibfnamefont {L.}~\bibnamefont {Pierre}},\ }\href
  {\doibase 10.1103/PhysRevLett.69.2590} {\bibfield  {journal} {\bibinfo
  {journal} {Phys. Rev. Lett.}\ }\textbf {\bibinfo {volume} {69}},\ \bibinfo
  {pages} {2590} (\bibinfo {year} {1992})}\BibitemShut {NoStop}%
\bibitem [{\citenamefont {Sindzingre}\ \emph {et~al.}(1994)\citenamefont
  {Sindzingre}, \citenamefont {Lecheminant},\ and\ \citenamefont
  {Lhuillier}}]{sindzingre}%
  \BibitemOpen
  \bibfield  {author} {\bibinfo {author} {\bibfnamefont {P.}~\bibnamefont
  {Sindzingre}}, \bibinfo {author} {\bibfnamefont {P.}~\bibnamefont
  {Lecheminant}}, \ and\ \bibinfo {author} {\bibfnamefont {C.}~\bibnamefont
  {Lhuillier}},\ }\href {\doibase 10.1103/PhysRevB.50.3108} {\bibfield
  {journal} {\bibinfo  {journal} {Phys. Rev. B}\ }\textbf {\bibinfo {volume}
  {50}},\ \bibinfo {pages} {3108} (\bibinfo {year} {1994})}\BibitemShut
  {NoStop}%
\bibitem [{\citenamefont {Yunoki}\ and\ \citenamefont
  {Sorella}(2006)}]{yunoki}%
  \BibitemOpen
  \bibfield  {author} {\bibinfo {author} {\bibfnamefont {S.}~\bibnamefont
  {Yunoki}}\ and\ \bibinfo {author} {\bibfnamefont {S.}~\bibnamefont
  {Sorella}},\ }\href {\doibase 10.1103/PhysRevB.74.014408} {\bibfield
  {journal} {\bibinfo  {journal} {Phys. Rev. B}\ }\textbf {\bibinfo {volume}
  {74}},\ \bibinfo {pages} {014408} (\bibinfo {year} {2006})}\BibitemShut
  {NoStop}%
\bibitem [{\citenamefont {Capriotti}\ \emph {et~al.}(1999)\citenamefont
  {Capriotti}, \citenamefont {Trumper},\ and\ \citenamefont
  {Sorella}}]{capriotti}%
  \BibitemOpen
  \bibfield  {author} {\bibinfo {author} {\bibfnamefont {L.}~\bibnamefont
  {Capriotti}}, \bibinfo {author} {\bibfnamefont {A.~E.}\ \bibnamefont
  {Trumper}}, \ and\ \bibinfo {author} {\bibfnamefont {S.}~\bibnamefont
  {Sorella}},\ }\href {\doibase 10.1103/PhysRevLett.82.3899} {\bibfield
  {journal} {\bibinfo  {journal} {Phys. Rev. Lett.}\ }\textbf {\bibinfo
  {volume} {82}},\ \bibinfo {pages} {3899} (\bibinfo {year}
  {1999})}\BibitemShut {NoStop}%
\bibitem [{\citenamefont {Weihong}\ \emph {et~al.}(1999)\citenamefont
  {Weihong}, \citenamefont {McKenzie},\ and\ \citenamefont
  {Singh}}]{zheng1999}%
  \BibitemOpen
  \bibfield  {author} {\bibinfo {author} {\bibfnamefont {Z.}~\bibnamefont
  {Weihong}}, \bibinfo {author} {\bibfnamefont {R.~H.}\ \bibnamefont
  {McKenzie}}, \ and\ \bibinfo {author} {\bibfnamefont {R.~R.~P.}\ \bibnamefont
  {Singh}},\ }\href {\doibase 10.1103/PhysRevB.59.14367} {\bibfield  {journal}
  {\bibinfo  {journal} {Phys. Rev. B}\ }\textbf {\bibinfo {volume} {59}},\
  \bibinfo {pages} {14367} (\bibinfo {year} {1999})}\BibitemShut {NoStop}%
\bibitem [{\citenamefont {White}\ and\ \citenamefont
  {Chernyshev}(2007)}]{white07}%
  \BibitemOpen
  \bibfield  {author} {\bibinfo {author} {\bibfnamefont {S.~R.}\ \bibnamefont
  {White}}\ and\ \bibinfo {author} {\bibfnamefont {A.~L.}\ \bibnamefont
  {Chernyshev}},\ }\href {\doibase 10.1103/PhysRevLett.99.127004} {\bibfield
  {journal} {\bibinfo  {journal} {Phys. Rev. Lett.}\ }\textbf {\bibinfo
  {volume} {99}},\ \bibinfo {pages} {127004} (\bibinfo {year}
  {2007})}\BibitemShut {NoStop}%
\bibitem [{\citenamefont {Miyake}(1992)}]{miyake1992}%
  \BibitemOpen
  \bibfield  {author} {\bibinfo {author} {\bibfnamefont {S.~J.}\ \bibnamefont
  {Miyake}},\ }\href {\doibase 10.1143/JPSJ.61.983} {\bibfield  {journal}
  {\bibinfo  {journal} {J. Phys. Soc. Jpn.}\ }\textbf {\bibinfo {volume}
  {61}},\ \bibinfo {pages} {983} (\bibinfo {year} {1992})}\BibitemShut
  {NoStop}%
\bibitem [{\citenamefont {Chubukov}\ \emph {et~al.}(1994)\citenamefont
  {Chubukov}, \citenamefont {Sachdev},\ and\ \citenamefont
  {Senthil}}]{chubukov1994}%
  \BibitemOpen
  \bibfield  {author} {\bibinfo {author} {\bibfnamefont {A.~V.}\ \bibnamefont
  {Chubukov}}, \bibinfo {author} {\bibfnamefont {S.}~\bibnamefont {Sachdev}}, \
  and\ \bibinfo {author} {\bibfnamefont {T.}~\bibnamefont {Senthil}},\
  }\href@noop {} {\bibfield  {journal} {\bibinfo  {journal} {J. Phys.: Cond.
  Mat.}\ }\textbf {\bibinfo {volume} {6}},\ \bibinfo {pages} {8891} (\bibinfo
  {year} {1994})}\BibitemShut {NoStop}%
\bibitem [{\citenamefont {Chernyshev}\ and\ \citenamefont
  {Zhitomirsky}(2009)}]{chernyshev2009}%
  \BibitemOpen
  \bibfield  {author} {\bibinfo {author} {\bibfnamefont {A.~L.}\ \bibnamefont
  {Chernyshev}}\ and\ \bibinfo {author} {\bibfnamefont {M.~E.}\ \bibnamefont
  {Zhitomirsky}},\ }\href {\doibase 10.1103/PhysRevB.79.144416} {\bibfield
  {journal} {\bibinfo  {journal} {Phys. Rev. B}\ }\textbf {\bibinfo {volume}
  {79}},\ \bibinfo {pages} {144416} (\bibinfo {year} {2009})}\BibitemShut
  {NoStop}%
\bibitem [{\citenamefont {Sachdev}(1992)}]{sachdev1992}%
  \BibitemOpen
  \bibfield  {author} {\bibinfo {author} {\bibfnamefont {S.}~\bibnamefont
  {Sachdev}},\ }\href {\doibase 10.1103/PhysRevB.45.12377} {\bibfield
  {journal} {\bibinfo  {journal} {Phys. Rev. B}\ }\textbf {\bibinfo {volume}
  {45}},\ \bibinfo {pages} {12377} (\bibinfo {year} {1992})}\BibitemShut
  {NoStop}%
\bibitem [{\citenamefont {Reuther}\ and\ \citenamefont
  {Thomale}(2011)}]{thomale2011}%
  \BibitemOpen
  \bibfield  {author} {\bibinfo {author} {\bibfnamefont {J.}~\bibnamefont
  {Reuther}}\ and\ \bibinfo {author} {\bibfnamefont {R.}~\bibnamefont
  {Thomale}},\ }\href {\doibase 10.1103/PhysRevB.83.024402} {\bibfield
  {journal} {\bibinfo  {journal} {Phys. Rev. B}\ }\textbf {\bibinfo {volume}
  {83}},\ \bibinfo {pages} {024402} (\bibinfo {year} {2011})}\BibitemShut
  {NoStop}%
\bibitem [{\citenamefont {Gvozdikova}\ \emph {et~al.}(2011)\citenamefont
  {Gvozdikova}, \citenamefont {Melchy},\ and\ \citenamefont
  {Zhitomirsky}}]{misha}%
  \BibitemOpen
  \bibfield  {author} {\bibinfo {author} {\bibfnamefont {M.~V.}\ \bibnamefont
  {Gvozdikova}}, \bibinfo {author} {\bibfnamefont {P.-E.}\ \bibnamefont
  {Melchy}}, \ and\ \bibinfo {author} {\bibfnamefont {M.~E.}\ \bibnamefont
  {Zhitomirsky}},\ }\href {\doibase 10.1088/0953-8984/23/16/164209} {\bibfield
  {journal} {\bibinfo  {journal} {J. Phys: Cond. Matt.}\ }\textbf {\bibinfo
  {volume} {23}},\ \bibinfo {pages} {164209} (\bibinfo {year}
  {2011})}\BibitemShut {NoStop}%
\bibitem [{\citenamefont {Seabra}\ \emph {et~al.}(2011)\citenamefont {Seabra},
  \citenamefont {Momoi}, \citenamefont {Sindzingre},\ and\ \citenamefont
  {Shannon}}]{seabra}%
  \BibitemOpen
  \bibfield  {author} {\bibinfo {author} {\bibfnamefont {L.}~\bibnamefont
  {Seabra}}, \bibinfo {author} {\bibfnamefont {T.}~\bibnamefont {Momoi}},
  \bibinfo {author} {\bibfnamefont {P.}~\bibnamefont {Sindzingre}}, \ and\
  \bibinfo {author} {\bibfnamefont {N.}~\bibnamefont {Shannon}},\ }\href
  {\doibase 10.1103/PhysRevB.84.214418} {\bibfield  {journal} {\bibinfo
  {journal} {Phys. Rev. B}\ }\textbf {\bibinfo {volume} {84}},\ \bibinfo
  {pages} {214418} (\bibinfo {year} {2011})}\BibitemShut {NoStop}%
\bibitem [{\citenamefont {Azaria}\ \emph {et~al.}(1992)\citenamefont {Azaria},
  \citenamefont {Delamotte},\ and\ \citenamefont {Mouhanna}}]{azaria92}%
  \BibitemOpen
  \bibfield  {author} {\bibinfo {author} {\bibfnamefont {P.}~\bibnamefont
  {Azaria}}, \bibinfo {author} {\bibfnamefont {B.}~\bibnamefont {Delamotte}}, \
  and\ \bibinfo {author} {\bibfnamefont {D.}~\bibnamefont {Mouhanna}},\ }\href
  {\doibase 10.1103/PhysRevLett.68.1762} {\bibfield  {journal} {\bibinfo
  {journal} {Phys. Rev. Lett.}\ }\textbf {\bibinfo {volume} {68}},\ \bibinfo
  {pages} {1762} (\bibinfo {year} {1992})}\BibitemShut {NoStop}%
\bibitem [{\citenamefont {Southern}\ and\ \citenamefont
  {Young}(1993)}]{southern93}%
  \BibitemOpen
  \bibfield  {author} {\bibinfo {author} {\bibfnamefont {B.~W.}\ \bibnamefont
  {Southern}}\ and\ \bibinfo {author} {\bibfnamefont {A.~P.}\ \bibnamefont
  {Young}},\ }\href {\doibase 10.1103/PhysRevB.48.13170} {\bibfield  {journal}
  {\bibinfo  {journal} {Phys. Rev. B}\ }\textbf {\bibinfo {volume} {48}},\
  \bibinfo {pages} {13170} (\bibinfo {year} {1993})}\BibitemShut {NoStop}%
\bibitem [{\citenamefont {Elstner}\ \emph {et~al.}(1993)\citenamefont
  {Elstner}, \citenamefont {Singh},\ and\ \citenamefont {Young}}]{singh}%
  \BibitemOpen
  \bibfield  {author} {\bibinfo {author} {\bibfnamefont {N.}~\bibnamefont
  {Elstner}}, \bibinfo {author} {\bibfnamefont {R.~R.~P.}\ \bibnamefont
  {Singh}}, \ and\ \bibinfo {author} {\bibfnamefont {A.~P.}\ \bibnamefont
  {Young}},\ }\href {\doibase 10.1103/PhysRevLett.71.1629} {\bibfield
  {journal} {\bibinfo  {journal} {Phys. Rev. Lett.}\ }\textbf {\bibinfo
  {volume} {71}},\ \bibinfo {pages} {1629} (\bibinfo {year}
  {1993})}\BibitemShut {NoStop}%
\bibitem [{\citenamefont {Zheng}\ \emph {et~al.}(2006)\citenamefont {Zheng},
  \citenamefont {Fj\ae{}restad}, \citenamefont {Singh}, \citenamefont
  {McKenzie},\ and\ \citenamefont {Coldea}}]{zheng2006}%
  \BibitemOpen
  \bibfield  {author} {\bibinfo {author} {\bibfnamefont {W.}~\bibnamefont
  {Zheng}}, \bibinfo {author} {\bibfnamefont {J.~O.}\ \bibnamefont
  {Fj\ae{}restad}}, \bibinfo {author} {\bibfnamefont {R.~R.~P.}\ \bibnamefont
  {Singh}}, \bibinfo {author} {\bibfnamefont {R.~H.}\ \bibnamefont {McKenzie}},
  \ and\ \bibinfo {author} {\bibfnamefont {R.}~\bibnamefont {Coldea}},\ }\href
  {\doibase 10.1103/PhysRevB.74.224420} {\bibfield  {journal} {\bibinfo
  {journal} {Phys. Rev. B}\ }\textbf {\bibinfo {volume} {74}},\ \bibinfo
  {pages} {224420} (\bibinfo {year} {2006})}\BibitemShut {NoStop}%
\bibitem [{\citenamefont {{L.G. Molinari}}\ and\ \citenamefont {{N.
  Manini}}(2006)}]{molinari}%
  \BibitemOpen
  \bibfield  {author} {\bibinfo {author} {\bibnamefont {{L.G. Molinari}}}\ and\
  \bibinfo {author} {\bibnamefont {{N. Manini}}},\ }\href {\doibase
  10.1140/epjb/e2006-00223-9} {\bibfield  {journal} {\bibinfo  {journal} {Eur.
  Phys. J. B}\ }\textbf {\bibinfo {volume} {51}},\ \bibinfo {pages} {331}
  (\bibinfo {year} {2006})}\BibitemShut {NoStop}%
\end{thebibliography}%

%\end{multicols}{2}
\end{document}